\documentclass[conference]{IEEEtran}

\ifCLASSINFOpdf
\else
\fi
\usepackage{cite}
\usepackage[nolist,nohyperlinks]{acronym}
\usepackage{comment}
\usepackage{tikz}
\usepackage{graphicx}
\usepackage{textcomp}
\usepackage[utf8]{inputenc}
\usepackage{multirow,makecell}
\usepackage[edges]{forest} 
\usetikzlibrary{shapes,trees} 
\usepackage{booktabs, siunitx} 
\usepackage[capitalize]{cleveref}
\usetikzlibrary{calc, positioning} 
\usetikzlibrary{intersections} 
\definecolor{blue1}{RGB}{66,146,198}
\definecolor{blue2}{RGB}{8,81,156}
\definecolor{blue3}{RGB}{8,29,88}
\definecolor{grey1}{RGB}{115,115,115}
\definecolor{grey2}{RGB}{82,82,82}
\definecolor{grey3}{RGB}{17,17,17}
\definecolor{red1}{RGB}{239,59,44}
\definecolor{green1}{RGB}{35,132,67}
\definecolor{green2}{RGB}{0,68,27}
\definecolor{teal1}{RGB}{53,151,143}
\definecolor{green3}{RGB}{53,151,143}
\definecolor{orange1}{RGB}{253,174,97}
\definecolor{orange2}{RGB}{204,76,2}
\usepackage[colorinlistoftodos]{todonotes}
\presetkeys{todonotes}{inline}{}

\begin{document}
\begin{acronym}

\acro{5GS}{5G system}
\acro{BMCA}{best master clock algorithm}
\acro{PTP}{precision time protocol}
\acro{gPTP}{generic PTP}
\acro{GM}{grandmaster}

\acro{PTP}{precision time protocol }

\acro{TAS}{time aware shaper}
\acro{TSN}{time-sensitive networking}
\acro{URLLC}{ultra-reliable low latency communication}
\acro{ITU-T}{international telecommunication union - telecommunication sector} 
\acro{GNSS}{global navigation satellite system}
\acro{gNB}{gNodeB}
\acro{RRH}{radio resource head}
\acro{PDU}{protocol data unit}
\acro{TT}{TSN translator}
\acro{UE}{user equipment}
\acro{UPF}{user plane function}

\acro{NW-TT}{network-side TT}
\acro{DS-TT}{device-side TT}
\acro{TSi}{ingress timestamp}
\acro{TSe}{egress timestamp}
\end{acronym}
\title{Time synchronization for deterministic communication}

 \author{\IEEEauthorblockN{Mahin K. Atiq and Raheeb Muzaffar\\
 \IEEEauthorblockA{Silicon Austria Labs GmbH, 4040 Linz, Austria, {firstname.lastname@silicon-austria.com}}}
}

\maketitle


\IEEEpeerreviewmaketitle

\section{Extended Abstract}
Deterministic communication is required for applications of several industry verticals including manufacturing, automotive, financial, and health care, etc. These applications rely on reliable and time-synchronized delivery of information among the communicating devices. Therefore, large delay variations in packet delivery or inaccuracies in time synchronization cannot be tolerated. In particular, the industrial revolution on digitization, connectivity of digital and physical systems, and flexible production design requires deterministic and time-synchronized communication. The IEC/IEEE~60802 \ac{TSN} profile describes traffic types, communication requirements, and message periodicity for several industrial applications.

A network supporting deterministic communication guarantees data delivery in a specified time with high reliability. The IEEE~802.1 \ac{TSN} task group is developing standards to provide deterministic communication through IEEE~802 networks. The \ac{TSN} technology is mainly based on four key features of time synchronization, traffic shaping and scheduling, reliability, and resource management. An interworking of a set of \ac{TSN} standards allow realization of these features for deterministic communication. Frequent time synchronization between the communicating devices is important due to clock drift that can accumulate significant time deviations~\cite{patel2021time}. The IEEE 802.1AS standard defines time synchronization mechanism for accurate distribution of time among the communicating devices. The time synchronization accuracy depends on accurate calculation of the residence time which is the time between the ingress and the egress ports of the bridge and includes the processing, queuing, transmission, and link latency of the timing information. Apart from time synchronization, traffic scheduling and mechanisms for reliable communication are also needed to realize deterministic communication. Traffic scheduling mechanisms like IEEE~802.1Qbv and  frame replication and elimination IEEE~802.1CB for reliability are defined in \ac{TSN}.

\textbf{Time synchronization in TSN:} The IEEE 802.1AS standard outlines the process for establishing a common sense of time across all devices connected to a TSN network, such as switches, bridges, and end stations~\cite{stanton2018distributing}. This shared sense of time is critical for ensuring the correct operation of other TSN mechanisms, including \ac{TAS}, as well as for coordinated functioning of different end stations. To achieve accurate time synchronization, the standard employs a \ac{gPTP} profile, which is based on the IEEE 1588v2 \ac{PTP}. In \ac{PTP}, the most accurate time source is selected as the \ac{GM} using the \ac{BMCA} and provides time reference to other devices in the network. \ac{PTP} sends sync. and follow-up messages from the time transmitter to the time receiver ports, synchronizing the local clocks of networked devices with each other. The standard also introduces redundancy in \ac{GM} devices and supports multiple \ac{gPTP} domains to ensure more stringent synchronization and enable the use of a redundant domain in case the performance of the first domain degrades. 

\textbf{Hot Standby Amendment:} Maintaining continuous time synchronization is crucial for ensuring deterministic communications, even in the event of failures. When the \ac{GM} fails in the IEEE802.1AS system without the hot standby, the \ac{BMCA} takes over to find a new \ac{GM}, leaving the network without synchronized time until a replacement is chosen. Additionally, the IEEE 802.1AS is only able to provide basic level of redundancy, i.e., the transient faults in the \ac{GM} (e.g., time glitches, excess jitter, and wander etc.) are not detected. This leads to the \ac{BMCA} flapping between potential \acp{GM} continuously. On the one hand, to reduce this waiting period and ensure resilient timing, the IEEE P802.1ASdm hot standby amendment aims to maintain two time domains simultaneously without relying on the \ac{BMCA}. The primary \ac{GM} and the hot standby \ac{GM} are chosen statically by an external entity (e.g., a remote management client). The hot standby \ac{GM} does not start transmitting timing information unless fully synchronized to the primary \ac{GM}. In this way, each end device (time receiver) will have two continuous time domains at all times and the failure mitigation is expected and predictable. 
On the other hand, the IEEE P802.1ASdm also detects transient faults in the \ac{GM} quality using the \emph{isSynced} function. Additionally, the hot standby amendment also allows for a \ac{GM} to have a arbitrary timescale. 
\begin{figure*}[!ht]
\centering
\resizebox{0.85\textwidth}{!}{
\input{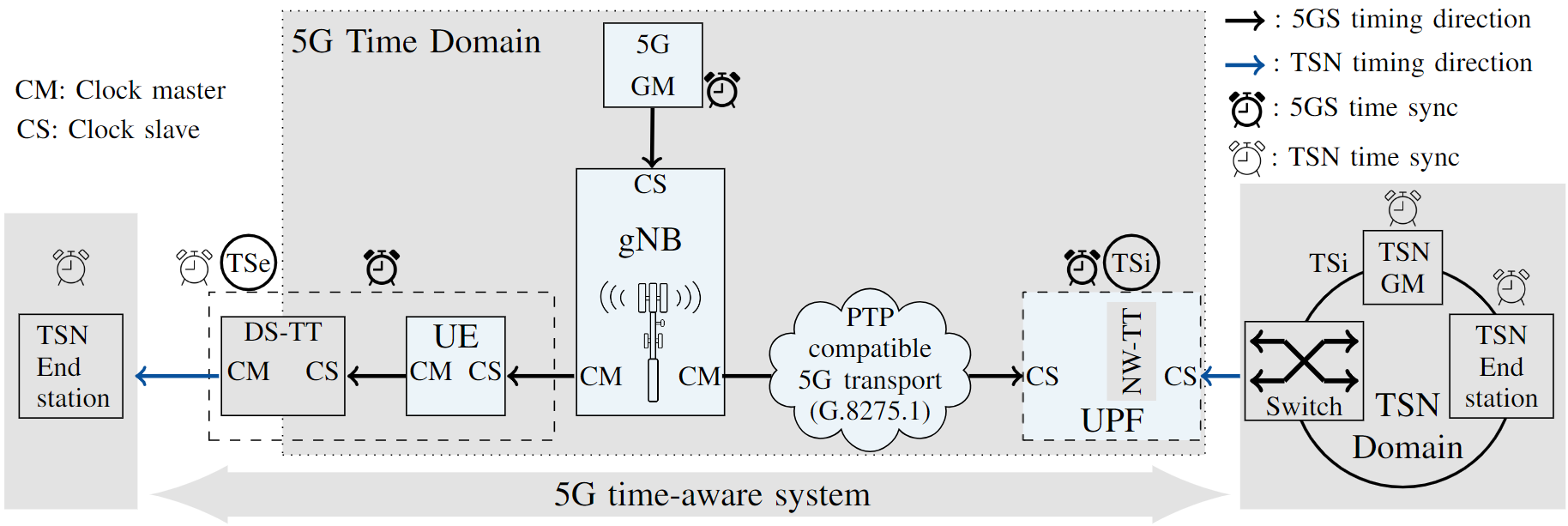}}
\caption{5G time-aware system supporting synchronization through boundary clock or transparent clock solutions~\cite{3GPP23501_17, 9652097}.}
\label{fig:5G_Time_Aware}
\end{figure*}

\textbf{Time synchronization in 5G:} The IEEE 802.1AS IEEE P802.1ASdm aims on time synchronization mechanisms in wired (Ethernet) network. Time synchronization is also essential in cellular networks for its operations. The time distribution mechanism for 5G is specified in G.8275.1 and G.8275.2 \ac{ITU-T} profiles and is also based on IEEE~1588 \ac{PTP} standard. The 5G baseband unit receives timing information from \ac{GNSS} as the source clock. The fronthaul synchronization between the \ac{RRH} and the baseband unit is serviced via enhanced common public radio interface. The \ac{gNB} synchronizes the \acp{UE} through exchange of \ac{PTP} messages~\cite{godor2020look}. A synchronization accuracy between $470$\,ns and $540$\,ns is achievable between the \ac{UE} and \ac{gNB} with $15$\,kHz sub-carrier spacing. Enhancements in 3GPP release-17 have been specified to reduce time errors incurred due to propagation delays between the \ac{gNB} and \acp{UE}.

\textbf{5G-TSN time synchronization:} The 5G communication technology also extends support for wireless deterministic communication through ultra-reliable low latency communication service and integration with \ac{TSN}. 
The \ac{5GS} integrates with \ac{TSN} as a logical bridge~\cite{3GPP23501_17}. The \ac{5GS} components including the radio access network and the core are synchronized using its internal clock (GNSS being the source clock). The wireless \ac{TSN} end stations synchronize with the \ac{TSN} domain by exchange of \ac{gPTP} messages through the 5G network~\cite{godor2020look}. 

The \ac{5GS} being a logical bridge provides connectivity between the \ac{TSN} domain and the \ac{TSN} end stations by establishing  \ac{PDU} sessions. The \ac{TT} functions namely the \ac{NW-TT} at the \ac{UPF} and \ac{DS-TT} at the \ac{UE} supports with \ac{gPTP} packet handling and timestamping (\cref{fig:5G_Time_Aware}). A \ac{gPTP} message sent from the \ac{TSN} domain to the \ac{TSN} end station is timestamped by the \ac{NW-TT} as an \ac{TSi} based on the \ac{5GS} internal clock. Similarly, the \ac{gPTP} message is timestamped by the \ac{DS-TT} with \ac{TSe} on exit. The residence time is calculated as the difference between the \ac{TSe} and \ac{TSi}. The residence time is added to the \ac{TSN} domain generated clock time and forwarded to the \ac{TSN} end stations for synchronization.  

The \ac{5GS} supports time synchronization service in one of the several modes including boundary clock, peer-to-peer transparent clock, end-to-end transparent clock, or as a time-aware system. Therefore, the clock source can be at the wired \ac{TSN} domain for downlink synchronization or at an end station for uplink synchronization in case of the transparent clock whereby the \ac{5GS} internal clock can be configured as the source to operate as a time-aware system or boundary clock. In addition, \ac{5GS} supported time synchronization service can be activated or deactivated by the \ac{TSN} application function. Moreover, to support timing resiliency to 5G applications sensitive to degradation in timing synchronization (examples include power grid and financial sector), Rel-18 (TS 22.261) provides mechanisms for 5G time resiliency. In this timing architecture, the 5G system acts as redundant or an alternative clock source to the time sensitive application.

\section{Conclusion}
\label{sec:conc}
Deterministic communication is required for several industrial applications where time awareness among the communicating devices is the key to configure deterministic operations. This article explains time synchronization mechanisms supported in current wired and wireless integrated systems. The latest amendments on IEEE P802.1ASdm hot standby and developments on 5G-TSN integration in 3GPP  are discussed. Further integration of hot standby with 5G/6G systems for time synchronization of end stations without significant discontinuity in case of a source clock failure is to be investigated.
\section*{Acknowledgment}
This work was supported by the European Union’s Horizon 2020 research and innovation programme through DETERMINISTIC6G project under Grant Agreement no. 101096504.

\bibliographystyle{IEEEtran}

\bibliography{Ref}

\end{document}